\documentclass[prb,twocolumn,longbibliography,amsmath,superscriptaddress,amssymb,nobibnotes]{revtex4-2}

\usepackage{graphicx}
\usepackage{dcolumn}
\usepackage{bm}
\usepackage{amssymb}
\usepackage{amsmath}
\usepackage{color}
\usepackage[dvipsnames]{xcolor}
\usepackage{placeins}
\usepackage{epstopdf}
\usepackage{multirow}
\usepackage{wasysym}

\usepackage[normalem]{ulem}

\usepackage[linkcolor=blue,urlcolor=blue,citecolor=blue,colorlinks=true]{hyperref}
\usepackage{makecell}

\begin{document}


\title{
Towards a critical endpoint in the valence fluctuating Eu(Rh$_{1-x}$Co$_{x}$)$_2$Si$_2$ system} 
\author{Franziska~Walther}
\email[]{franziska.walther@stud.uni-frankfurt.de}
\affiliation{%
 Kristall- und Materiallabor, Physikalisches Institut, 
 Goethe-Universit\"at Frankfurt, Max-von-Laue Stasse 1, 
 60438 Frankfurt am Main, Germany
}%
\author{Michelle~Ocker}
\affiliation{%
 Kristall- und Materiallabor, Physikalisches Institut, 
 Goethe-Universit\"at Frankfurt, Max-von-Laue Stasse 1, 
 60438 Frankfurt am Main, Germany
}%
\author{Alexej Kraiker}
\affiliation{%
 Kristall- und Materiallabor, Physikalisches Institut, 
 Goethe-Universit\"at Frankfurt, Max-von-Laue Stasse 1, 
 60438 Frankfurt am Main, Germany
}%
\author{Nubia~Caroca-Canales}
\affiliation{Max Planck Institute for Chemical Physics of Solids, 01187 Dresden, Germany}
\author{Silvia Seiro}
\affiliation{Max Planck Institute for Chemical Physics of Solids, 01187 Dresden, Germany}
\affiliation{Leibniz Institute for Solid State and Materials Research Dresden, 01069 Dresden, Germany}
\author{Kristin~Kliemt}
\email[Corresponding author:]{kliemt@physik.uni-frankfurt.de}
\affiliation{%
 Kristall- und Materiallabor, Physikalisches Institut, 
 Goethe-Universit\"at Frankfurt, Max-von-Laue Stasse 1, 
 60438 Frankfurt am Main, Germany
}%
\author{Cornelius~Krellner}
\affiliation{%
 Kristall- und Materiallabor, Physikalisches Institut, 
 Goethe-Universit\"at Frankfurt, Max-von-Laue Stasse 1, 
 60438 Frankfurt am Main, Germany
}%

\date{\today}

\begin{abstract}
We report on the successful single crystal growth of pure EuRh${_2}$Si${_2}$ and of Eu(Rh$_{1-x}$Co$_{x}$)$_2$Si$_2$ with $x\leq0.23$ by the flux method. Through Co substitution, EuRh$_2$Si$_2$ can be tuned from stable antiferromagnetism via a valence-transition state towards the valence-crossover regime.  From magnetization measurements, we constructed a $B - T$ phase diagram for EuRh${_2}$Si${_2}$ comprising multiple magnetic phases and showing a sizable magnetic anisotropy within the basal plane of the tetragonal unit cell. This indicates a complex antiferromagnetic ground state for $x=0$. By applying positive chemical pressure through the substitution series Eu(Rh$_{1-x}$Co$_{x}$)$_2$Si$_2$, a sharp temperature-induced first-order phase transition is observed in magnetization, resistivity and heat capacity for 0.081 $\leq$ $x$ $\leq$ 0.119. The critical end point of this valence transition is located in the phase diagram in the vicinity of 0.119 $<x_{\rm EDX}<$ 0.166. At higher substitution level, the system reaches a valence-crossover regime. The obtained results are presented in a temperature-substitition phase diagram.
\end{abstract}

                             
\maketitle


\section{\label{sec:level1}Introduction}
In ternary europium-based intermetallic compounds a variety of intriguing electronic properties have been observed e.g.
skyrmions \cite{Kakihana2019,Sereni2023}, nontrivial topological phases \cite{riberolles2021magnetic}, colossal magnetoresistance \cite{Krebber2023,Souza2022,Mueller2014} and in particular valence instabilities \cite{dionicio2006temperature,mitsuda2000eu,segre1982valence}. 
While these intriguing phenomena originate from electronic ordering, recent years have seen a growing focus on harnessing the emergence of novel effects resulting from the strong coupling between electronic and lattice degrees of freedom.
An exemplary material that shows the emergence of a new form of elasticity due to electron-lattice coupling is the organic charge transfer salt $\kappa$-(BEDT-TTF)$_2$Cu[N(CN)$_2$]Cl, which reveals a pressure-induced first-order Mott-metal-insulator transition, which terminates at a critical end point (CEP) \cite{gati2016breakdown}. 
By probing the elastic response of the material as it passes through the CEP, critical elasticity emerges around the CEP, characterized as the breakdown of Hooke's law. In order to study the electron-lattice coupling in intermetallic systems close to a CEP, $4f$-systems such as YbInCu$_4$ \cite{ocker2025single} and Eu(Pd$_2$Si$_{1-x}$Ge$_{x}$)$_2$ \cite{peters2023valence,wolf2023pressure} have proven to be suitable systems, as they show first-order valence transitions accompanied by pronounced changes in the crystal lattice. \\ 
In Eu compounds crystallizing in the ThCr$_2$Si$_2$-type structure, the Eu ions can adopt a magnetic divalent state, which is associated with a large unit cell volume, as observed in the antiferromagnetic compound EuRh$_2$Si$_2$ \cite{seiro2013complex}.  
Alternatively, Eu can exist in a non-magnetic trivalent state with a smaller unit cell, as in EuCo$_2$Si$_2$, where only Van Vleck paramagnetism is present \cite{seiro2013anomalous}.
This is because the valence configurations of Eu$^{2+}$ and Eu$^{3+}$ are energetically rather close \cite{MIEDEMA1976167} and thus intermediate-valence states between a nearly divalent Eu$^{(2+\delta)+}, (\delta <1)$, and a nearly trivalent electronic state Eu$^{(3-\delta')+}, (\delta'<1)$ may emerge. These valence fluctuations can be driven by different external parameters such as temperature \cite{mimura2004bulk}, pressure  \cite{honda2017pressure, dionicio2006temperature}, magnetic field \cite{scherzberg1984field} and substitution \cite{wada1997temperature,fukuda2003application}. The general $p$-$T$ phase diagram for Eu compounds, crystallizing in the ThCr$_2$Si$_2$ type structure, reveals a stable divalent electronic state of Eu at low pressures (i.e. large unit-cell volume) and low temperatures with magnetic ordering. 
 With increasing pressure, the magnetic ordering disappears abruptly and the systems show a valence transition from Eu$^{2+}$ to Eu$^{(3-\delta')+}$ with temperature \cite{onuki2020unique}. These first-order phase transitions terminate in a critical end point (CEP) of second order, while the system enters a valence-crossover regime beyond the CEP at high pressures (small unit-cell volume). In proximity of the critical end point, the magneto-elastic coupling of the valence transition may lead to the occurrence of critical elasticity as observed in \cite{gati2016breakdown}. \\
 In order to study the electron-lattice coupling close to the critical end point using external pressure, a material system which is localized at the low-pressure side of the CEP has to be found.  By applying hydrostatic pressure, it is then possible to pass through the region of putative critical elasticity and probe the elastic response of the system. A suitable material that can be tuned by chemical pressure in the vicinity of the CEP is EuRh${_2}$Si${_2}$. Previous studies under hydrostatic pressure \cite{honda2016pressure} show a stable divalent state at pressures $p\leq 0.96$\,GPa and a valence transition realized between $0.96 \leq p \leq 2$\,GPa, which terminates at an estimated pressure of $p_{\rm CEP} = 2.05$\,GPa at the critical end point. \\
  To induce positive chemical pressure comparable to  hydrostatic pressure, an isoelectronic and isostructural substitution is required, e.g. substitution of Rh atoms by smaller Co atoms in the series Eu(Rh$_{1-x}$Co$_{x}$)$_2$Si$_2$.  
 Since EuCo${_2}$Si${_2}$ crystallizes in the same structure with a 11\% smaller unit cell \cite{mayer1972}, a large compression of the unit cell is expected, when Co is introduced into the lattice. Under the assumption of a linear decrease of the lattice parameters with $x$ and correspondingly a linear decrease of the unit-cell volume (Vegards law), the hydrostatic pressure can be calculated using a typical bulk modulus of EuRh${_2}$Si${_2}$ of $K$ = 100\,GPa \cite{onuki2020unique}. 
 Hence, the valence transition should take place for substitution levels between 0.08 $\leq$ $x$ $\leq$ 0.18 while the CEP is expected for $x^{\rm CEP}$ = 0.18. Compared to another substitution series Eu(Rh$_{1-x}$Ir$_{x}$)$_2$Si$_2$ \cite{seiro2011stable}, where the CEP is localized in the range of 0.5 $<$ $x$ $<$ 0.75, in the here presented system a lower substitution level is necessary to reach the CEP. This is important, as higher substitution concentrations can induce more disorder in the system, and thus strongly affect the behavior at the CEP. So far, the substitution series with Co has been investigated on polycrystalline samples for $x$ = 0.4 by high-energy resolution fluorescence detection X-ray absorption spectroscopy (HERFD-XAS) \cite{shimokasa2020electronic} and hard x-ray photoemission spectroscopy \cite{ichiki2017valence} as well as for $x = 0.1, 0.2; 0.3$ by x-ray absorption spectroscopy (XAS) \cite{ichiki2017valence}. In contrast to the pure system EuRh${_2}$Si${_2}$, which shows a temperature-independent Eu valence of $\sim$ 2.1 \cite{ichiki2017valence} for $x = 0.4$ a gradual increase of the Eu valence from 2.57 at 300\,K to 2.92\,K is observed \cite{ichiki2017valence} indicating the valence crossover. For lower substitution concentrations with $x = 0.1 $, a drastic valence change occurs from 2.2 at 300\,K to 2.5 at 18\,K, characteristic for a valence transition \cite{ichiki2017valence}. Hence, the substitution series Eu(Rh$_{1-x}$Co$_{x}$)$_2$Si$_2$ is a promising system for studying valence fluctuations on single crystals, which couples to the crystal lattice.\\  
 In this work, single crystals of the pure system EuRh${_2}$Si${_2}$ have been studied at first to examine their magnetic ground state by magnetization measurements with focus on the magnetic anisotropy within the basal plane of the crystal lattice. Chemical substitution may also affect magnetic anisotropy, as observed in EuPd$_2$(Si$_{1-x}$Ge$_{x}$)$_2$ \cite{peters2023valence} and thus a detailed understanding of the complex anisotropic magnetic behavior is required. Then, the growth of single crystals of the substituted system Eu(Rh$_{1-x}$Co$_{x}$)$_2$Si$_2$ for different substitution levels is reported, which will be characterized by measurements of magnetization, heat capacity, and resistivity.

\section{Experimental Details}
\subsection{Crystal Growth}\noindent
The single crystal growth of EuRh${_2}$Si${_2}$ as well as Eu(Rh$_{1-x}$Co$_{x}$)$_2$Si$_2$ proved challenging due to a high melting point of rhodium with $T{_m} = 1966\,^{\circ}$C, which is combined with a low boiling point of europium with $T{_b} = 1439\,^{\circ}$C and therefore a high vapor pressure above $1300\,^{\circ}$C. To reduce the high melting point of rhodium, the flux method is used. Single crystals of EuRh${_2}$Si${_2}$ were grown in indium flux using the Bridgman technique as described in \cite{seiro2013complex}. Whereas single crystals of Eu(Rh$_{1-x}$Co$_{x}$)$_2$Si$_2$ were grown in Eu-flux for different nominal substitution levels $0.10 \leq x \leq 0.30$ with two initial compositions 1.8:2:2 and 2.5:2:2 as summarized in table \ref{tab:overwiew}. On the first growth approaches, an additional prereaction of Rh$_{1-x}$Co$_{x}$ was performed in an argon arc furnace (see Tab.~\ref{tab:overwiew}). High-purity starting materials Eu (EvoChem, 99.99\%, pieces), Rh (EvoChem, 99,9\%, powder; $\>$ 99.9\%, pieces), Co (Chempur, 99.99\%, powder; Johnson Matthey Chemicals Limited, 99.9\%, rod), Si (Cerac, 99.9999\%, pieces) were placed in a screwable inner graphite crucible to avoid europium evaporation and to prevent a reaction of the highly reactive melt with the outer crucible material.\\ 
Depending on the initial mass of the melt ($m = 2$\,g and $m = 4$\,g), graphite crucibles with different dimensions were used. The smaller graphite crucible has an inner diameter of $\diameter_i = 8$\,mm and a length of $l = 53$\,mm, whereas the  larger graphite crucible has a diameter of  $\diameter_i = 13$\,mm and a length $l =  48$\,mm (see Tab.~\ref{tab:overwiew}). 
The inner crucible is sealed in a niobium or tantalum outer crucible. The growth was carried out in a vertical resistive furnace (GERO HTRV70-250/18) under an argon atmosphere. The crucible was  heated to $1520\,^{\circ}$C, homogenized for 5 hours, followed by a slow cooling period to $1000\,^{\circ}$C with 4\,K/h. Finally, the furnace was cooled to room temperature with fast rate of 150\,K/h.

\begin{table}[ht!]
    \centering
\caption{Overview of single crystalline samples with nominal Co-concentration $x_{\rm nom}$, incorporated Co-concentration $x_{\rm EDX}$ in the crystal determined by EDX-measurements, crucible size, stoichiometry and prereaction. }
\label{tab:overwiew}
    \begin{tabular}{|c|c|c|c|c|c|} \hline 
         $x_{\rm EDX}$&0.06& 0.08& 0.11& 0.12& 0.23\\ \hline 
 $x_{\rm nom}$& 0.10& 0.3& 0.2& 0.2& 0.3\\ \hline 
 crucible & small& large&large& small&small\\ \hline 
 \makecell{initial composition\\Eu:Rh$_{1-x}$Co$_{x}$:Si}& 1.8:2:2 & 2.5:2:2 &2.5:2:2 & 2.5:2:2&2.5:2:2\\ \hline
 \makecell{prereaction\\Rh$_{1-x}$Co$_{x}$:Si}& $\checkmark$& -&   -& $\checkmark$&-\\\hline
    \end{tabular}    
\end{table}

\noindent For structural characterization, powder X-ray diffraction (PXRD) was performed using a Bruker D8 diffractometer (Cu-K$_{\alpha}$ radiation with $\lambda =1.5406$\,\r{A} and Bragg-Brentano geometry). The chemical composition of the grown crystals, especially the substitution concentration $x_{\rm EDX}$ = Co/(Rh+Co) was analyzed by energy-dispersive x-ray spectroscopy (EDX) using a Zeiss-DSM 940A scanning electron microscope with an EDAX detector. To determine the orientation of the grown single crystals and their crystallinity, a M\"uller Micro Laue instrument was used with white X-rays from a tungsten anode. The commercial Quantum Design Physical Property Measurement System (PPMS) was utilized to measure magnetization, heat capacity, and resistivity. The latter was determined using the four-point alternating current transport (ACT) configuration. The heat capacity was measured with the thermal relaxation technique. A heating pulse is applied to the sample, afterwards the system slowly cools down and relaxes to thermal equilibrium. The heat capacity data was evaluated separately for the heating and cooling curves of a large heat pulse with a typical temperature rise of 30\,K.

\section{Results and Discussion}
\subsection{Structural and chemical characterization}\noindent

\begin{figure} [ht!]
\centering
\includegraphics[width=0.5\textwidth]{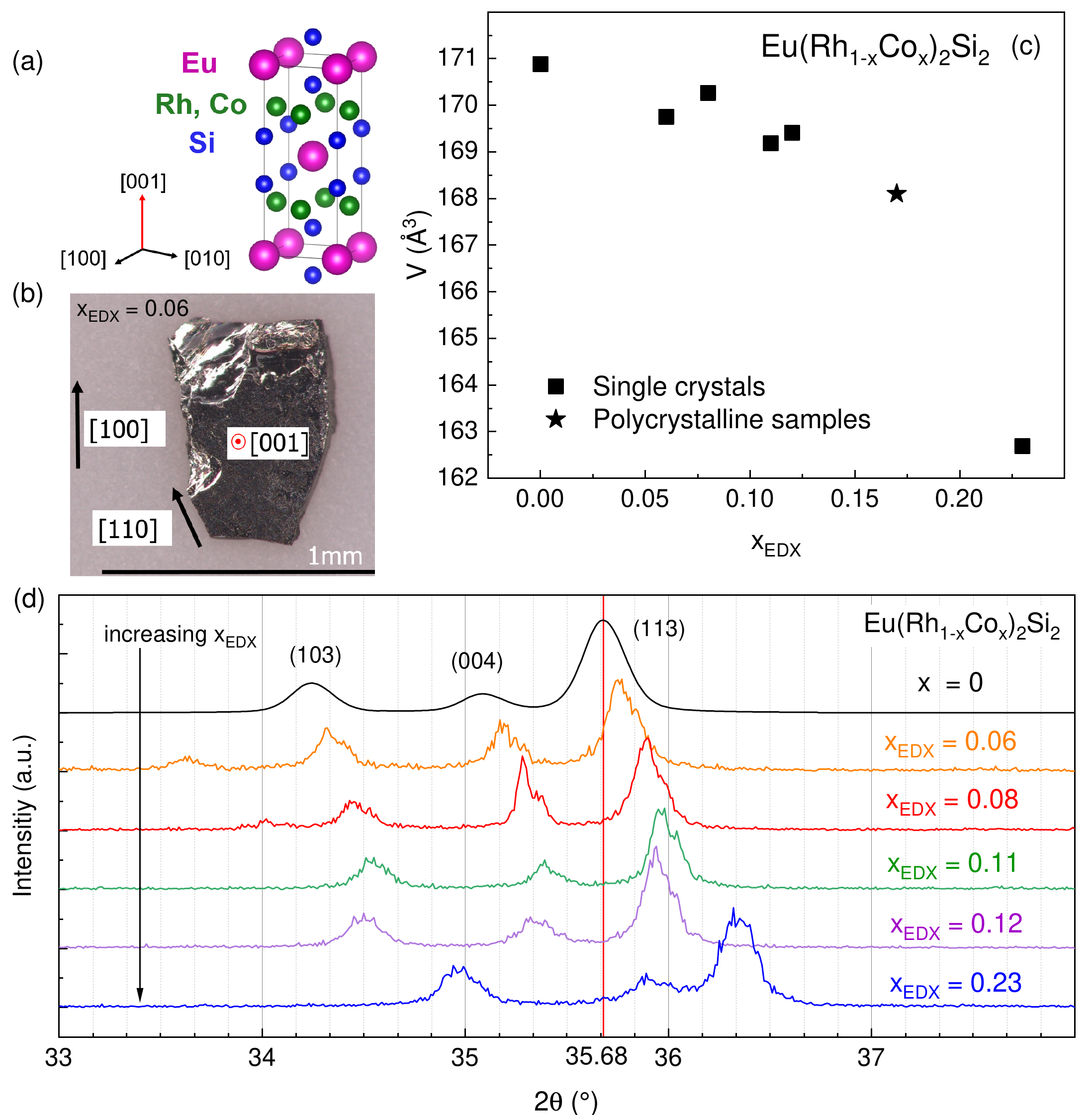}
	\caption{(a): ThCr${_2}$Si${_2}$-type tetragonal crystal structure, (b): Single crystal with $x_{\rm EDX}$ = 0.06, (c): Unit-cell volume as function of Co concentration (d): X-ray diffractogram of the substitution series Eu(Rh$_{1-x}$Co$_{x}$)$_2$Si$_2$ for selected reflexes. Black data for $X=0$ taken from \cite{Ballestracci}.}
\label{diffractogram}
\end{figure}

The crystal growth yields small substituted single crystals with $\leq$ 1\,mm in length (Fig.~\ref{diffractogram}b). The evaluation of the Laue images shows  that the thin plates are aligned in the $a-a$ plane. The chemical composition of the 122 phase was confirmed by EDX measurements, which revealed two slightly different compositions with tiny variations in the contents of Eu and Si. 
Even within a single sample, regions deficient in Eu and enriched in Si can be observed. The stoichiometry ranges from Eu:(Rh+Co):Si = ($16.9 \,\pm 0.30$) : ($39.9 \,\pm 1.2$): ($43.1 \,\pm 1.1$), where the denoted uncertainties are the standard deviations of the average value, to a Eu-rich phase with a lower proportion of Si with Eu:(Rh+Co): Si = ($19.7 \,\pm 1.60$) : ($41.2 \,\pm 0.4$): ($39 \,\pm 1.6$). 
The latter phase is closer to the expected 1:2:2 stoichiometry. The (Rh + Co) content is nearly constant in both phases. Furthermore, one has to mention that the Co concentration $x_{\rm EDX}$ in the crystal determined by EDX analysis is influenced by the Eu content, as in the EDX spectrum the Co-K line at $E = 6.915$\,keV overlaps with the Eu-L$_{\gamma1}$ line at $E = 6.891$\,keV. 
Due to these measurement uncertainties, the amount of Co can be underestimated for low Co concentrations, when the Co peak is not very pronounced and lies on the shoulder of the Eu-L$_{\gamma1}$ line. \\ In Tab.~\ref{tab:overwiew}, an overview of the growths performed for three different nominal substitution concentrations $x_{\rm nom}$ = 0.1; 0.2; 0.3 is given, where $x_{\rm EDX}$ is the average Co concentration detected in the crystals within one batch by EDX measurement. 
Different concentrations $x_{\rm EDX}$ of Co were found in different batches with the same nominal concentration. Thus, the incorporation rate $k_{Co}=x_{\rm nom}/x_{\rm EDX}$ depends on experimental parameters, such as the size of the crucible, mixing conditions, and whether or not a prereaction was performed. For a homogeneous incorporation of Co in the crystal, a sufficient mixing of the melt is necessary. Our results allow for the conclusion that a smaller crucible size allows for better mixing of the melt and leads to a higher incorporation of Co. \\
\noindent Powder x-ray diffraction confirmed the tetragonal structure with the $I4/mmm$ space group for all substitution concentrations. In Fig.~\ref{diffractogram}d, the diffractogram of Eu(Rh$_{1-x}$Co$_{x}$)$_2$Si$_2$ for different substitution concentrations is shown for selected reflexes. No significant broadening of the reflexes was detected, suggesting that there was no large variation of the Co concentration within the powder sample. 
For the high-intensity reflex (1\,1\,3) there is a shift to higher angles as the concentration $x_{\rm EDX}$ increases. This is due to the compression of the unit-cell volume, shown in Fig.~\ref{diffractogram}c, as expected by applying positive chemical pressure. 
The volume of the unit cell decreases linearly with increasing substitution concentration for low concentrations, obeying Vegards law, while a stronger  decrease is observed for higher substitution levels $x > 0.17$. This results in a large relative unit cell compression of 4.8\% for $x_{\rm EDX} = 0.23$.
\subsection{EuRh${_2}$Si${_2}$}
\subsubsection{Magnetization}
\noindent  In Fig.~\ref{fig:MvTcomparison}, the temperature-dependent magnetic susceptibility, $\chi(T)$, measured in zero field cooled (ZFC) and field cooled (FC) mode is compared for field aligned along the in-plane directions $[100]$ and $[110]$ for a small applied magnetic field of $\mu_0H = 0.005$\,T. 
The compound EuRh${_2}$Si${_2}$ undergoes three magnetic phase transitions below $T_{\rm N} = 24$\,K. The transition temperatures at $T_1$ and $T_2$ were determined by the kink in the susceptibility of the ZFC curve, while $T_1$ was taken as the midpoint of the rise. For the field applied along the $[100]$ direction (black curve), the first transition appears at $T_1 = 24.9$\,K followed by the second transition at $T_2 = 23.7$\,K and a possible first-order transition at $T_3 = 11.6$\,K as previously reported in reference \cite{seiro2013complex} (open symbols in Fig.~\ref{fig:MvTcomparison}).\\  
Below $T_3$, the ZFC and FC curves differ, which may originate from a reorientation of magnetic domains or a ferromagnetic component.
For field along the $[110]$ direction (blue curve), only slight differences are observed for the transition temperatures with $T_1 = 24.8$\,K, $T_2 = 23.4$\,K and $T_2 = 11.8$\,K. However, the susceptibility at low temperatures for field along $[100]$ is larger than for $[110]$. The observed negative susceptibility for the ZFC curve for field applied along the $[110]$ direction is attributable to the diamagnetic signal of the quartz sample holder and a larger amount of GE varnish. 

In the inset of Fig.~\ref{fig:MvTcomparison} (a), the inverse magnetic susceptibility, $\chi^{-1}(T)$, for the field of $\mu_0H = 1$\,T applied along the in-plane directions is shown, which exhibits Curie-Weiss behavior above $T_{\rm N}$. 
Data were fitted using the Curie-Weiss expression in the temperature range 150\,K$\leq T \leq 300$\,K. The obtained Curie-Weiss temperatures $\Theta_{100} = (27\,\pm 1)$\,K and $\Theta_{110} = (29\,\pm 1)$\,K as well as the effective magnetic moments $\mu^{100}_{\rm eff} = (7.7\pm 0.1)\mu_B$ and $\mu^{110}_{\rm eff} = (7.8\,\pm 0.1)\mu_B$ show no significant anisotropy between the in-plane directions as expected for a system with $L = 0$ and negligible crystal electric field effects. The Curie-Weiss temperature for a field applied along the $[100]$ direction is in accordance with \cite{seiro2013complex}.
The effective magnetic moment for both field directions is very close to the theoretically expected value for Eu$^{2+}$ with $\mu^{\rm calc}_{\rm eff}=7.94\,\mu_B$.\\
\noindent In Fig.~\ref{fig:MvTcomparison} (b), the magnetization as a function of the magnetic field, $M(H)$, at $T$ = 2\,K is shown for the in-plane and out-of-plane directions. 
The data recorded with the field applied along the $[001]$ direction reveal a continuous evolution with the magnetic field. However, a spin-flop-like transition with a sharp hysteresis appears below $\mu_0H=1\,\rm T$ for this field direction, where the origin is not clear yet. Since this pronounced hysteresis is reproducible in comparison to \cite{seiro2013complex}, we do not attribute the hysteresis to a misalignment of the sample. The data with field applied along an in-plane direction show  metamagnetic transitions at $B_1$, accompanied by a hysteresis, Fig.~\ref{fig:MvTcomparison}(c), as already observed in \cite{seiro2013complex} and the field-polarized state with 7\,$\mu_B$ per Eu atom is reached at $B_2$. A clear in-plane anisotropy for the two field directions $[100]$ and $[110]$ is observed below $B_2$.
The characteristic fields $B_1$ and $B_2$ were determined by the kink in the magnetization measured with increasing field.

In Fig.~\ref{fig:MvTcomparison}(d), the field-dependent magnetic susceptibility, $M$/$\mu_0H$ versus $\mu_0H$ is shown for different applied field directions to obtain information about the (re)orientation of magnetic moments for small magnetic fields, as demonstrated for the sister compound GdRh$_2$Si$_2$ in \cite{kliemt2017gdrh}. 
There, a characteristic field dependence of the susceptibility was observed, which is indicative of a flipping of magnetic domains. In the case of EuRh$_2$Si$_2$, the situation is more complex but the low-field data is reminicent of the GdRh$_2$Si$_2$ case.
The data show that a change in $M$/$\mu_0H$ occurs below $B^{100,*}=0.015\,\rm T$  and below $B^{110,*}=0.007\,\rm T$ which could be connected to a flipping of (a part of the) magnetic domains in the $a-a$ plane. 
Even a small field is enough to reorient the magnetic moments out of the magnetic ground state configuration. Thus, the ground state and the spin configurations above $B^*$ are energetically very close. 
Here, a weak magnetic in-plane anisotropy is observable. 
For the field applied along the $[001]$ direction (red curve in Fig.~\ref{fig:MvTcomparison}(d)), the field-dependent magnetic susceptibility shows a constant behavior for low magnetic fields and suggests a nearly linear evolution of the magnetization with field. For a simple antiferromagnetic system, this corresponds to the case when the magnetic field is applied perpendicular to the magnetic sublattices. Thus, one suggests that the magnetic moments are aligned in the basal plane perpendicular to the $[001]$ direction at very low fields. This is consistent with the proposed scenario of ferromagnetic layers stacked along $c$ with a propagation vector $(0, 0, \tau)$ \cite{chikina2014strong} with moment alignment in the $a-a$ plane observed through M\"ossbauer spectroscopy \cite{FELNER1984419}. It is possible that the magnetic moments are slightly tilted out of the basal plane, which is not resolvable in the $M$/$\mu_0H$ representation. For instance, a small parallel component along the $c$-axis could thus generate a spin-flop-like transition as the one observed below $\mu_0H^{001}=1\,\rm T$ (red curve in Fig.~\ref{fig:MvTcomparison}(b)).

For field applied along the in-plane directions, we observe an abrupt increase at $B_1$  which may be attributable to a reorientation of magnetic moments through a spin-flop-like transition.
For the field applied along the $[100]$ direction, this metamagnetic transition occurs at $B_1$ = 0.08\,T, Fig.~\ref{fig:MvTcomparison}(c), and the field-polarized state is reached at $B_2$ = 2.8\,T. 
However, for the $[110]$ direction, a similar metamagnetic transition occurs in a magnetic field of $B_1$ = 0.07\,T. Above $B_1$, the magnetization increases smoother than in the case of a field applied along the $[100]$ direction, and the field-polarized state is thus reached at a higher field of $B_2$ = 5.4\,T. 
The particular behavior shown in Fig.~\ref{fig:MvTcomparison}(c) reveals that the situation is complex as the spin-flop character occurs for both applied field directions. Furthermore, from the field dependence of the susceptibility, $M(H)$/$H$, of the inplane directions, one can deduce that the magnetic moments are arranged in a non-collinear structure in the basal plane, e.g. spiral or fan structure, which would be in line with the proposed incommensurable structure proposed in \cite{seiro2013complex}. This hints at an alignment of the magnetic moments along a direction other than $[100]$ or $[110]$ for fields lower than $B_1$ maybe in a further fan or spiral structure.

\subsubsection{Phase diagram}
\begin{figure}[ht!]
    \centering
    \includegraphics[width=1\linewidth]{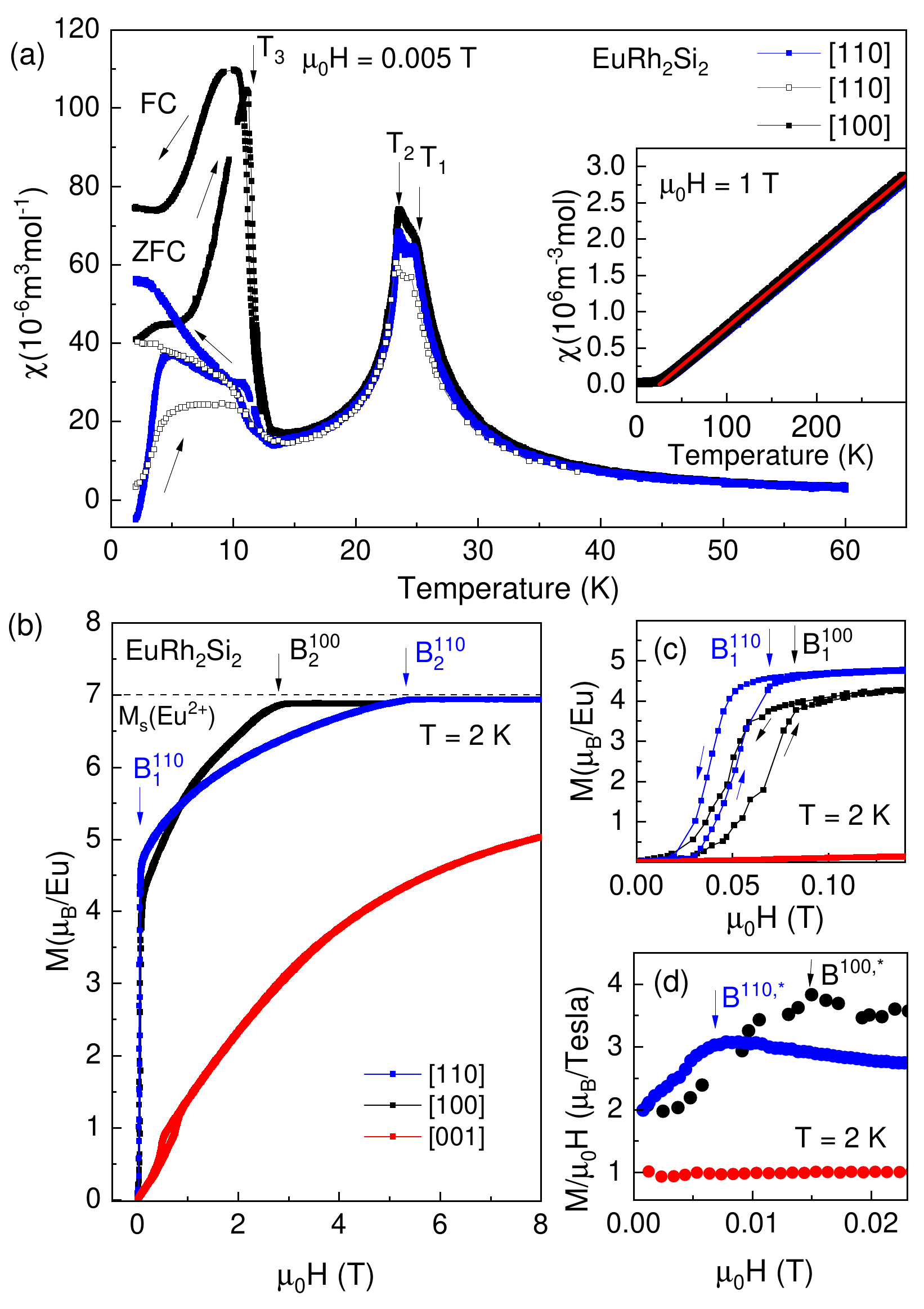}
    \caption{EuRh$_2$Si$_2$: (a) Magnetic susceptibility as a function of temperature for field applied along the $[100]$ and $[110]$ directions at $\mu_0H$ = 0.005\,T. Data with open symbols are taken from \cite{seiro2013complex}. Inset presents the temperature dependent inverse susceptibility at $\mu_0H$ = 1\,T for fields applied along the in-plane directions $[100]$ and $[110]$. (b) Magnetization as a function of magnetic field for the in-plane and out-of-plane directions at $T = 2\,\rm K$. (c) Spin-flop like transitions for the two in-plane directions at $T = 2\,\rm K$.  (d) $M$/$\mu_0H$ versus $\mu_0H$ for small magnetic field at $T = 2\,\rm K$. }
    \label{fig:MvTcomparison}
\end{figure}

\begin{figure}[ht!]
    \centering
    \includegraphics[width=1\linewidth]{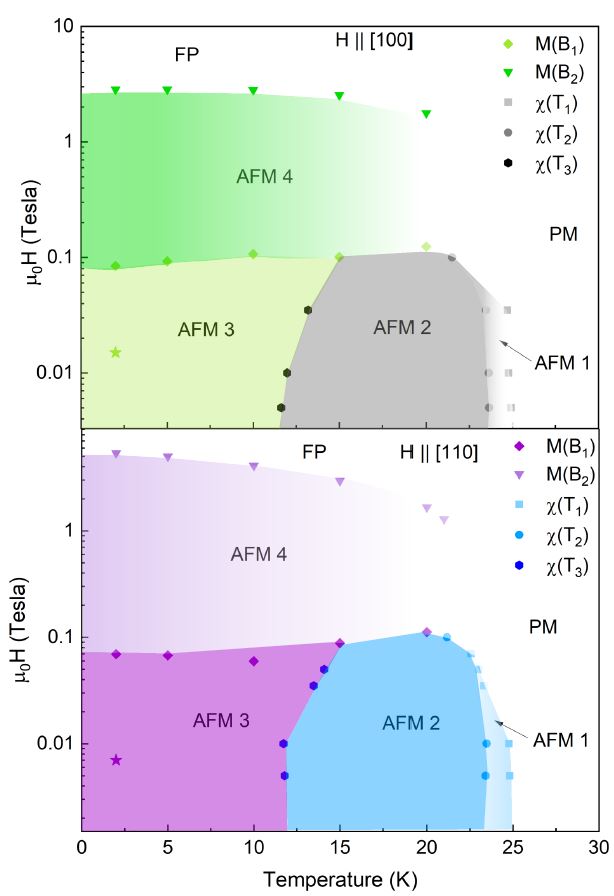}
    \caption{EuRh$_2$Si$_2$: Magnetic phase diagram for field applied along $[100]$ (top) and $[110]$ (bottom).}
    \label{fig:PD_100}
\end{figure}

\noindent From the magnetic susceptibility and magnetization data,  Fig.~\ref{fig:MvTcomparison} (data for higher fields and temperatures not shown), one can construct a field-temperature phase diagram for the $[100]$ and $[110]$ directions, which is shown in Fig~\ref{fig:PD_100}. 
The $B-T$ phase diagram indicates four antiferromagnetic phases showing a complex magnetic structure. Below the Néel temperature $T_{\rm N} = 24$\,K, EuRh${_2}$Si${_2}$ enters the AFM 1 phase. With decreasing temperature, the  AFM 2 phase appears at $T_2$, which slightly shifts to lower temperatures with increasing field. Further cooling forces a possible first-order transition at $T_3$ into the low-temperature AFM 3 phase (fan / spiral structure with moments in plane). This phase stabilizes to higher temperatures with an increasing magnetic field. For small magnetic fields within the AFM 3 regime, the star symbol is assigned to the observed critical magnetic field $B^*$ in the $M$/$\mu_0H$ versus $\mu_0H$ representation.
We suspect another magnetic phase (with fan / spiral alignment of the moments in the $a-a$ plane) at low temperatures and small fields below $B^*$, which represents the ground state of the material. 
However, the phase boundary lines are not yet clear. 
For magnetic fields larger than $B_1$, the system moves into the AFM 4 phase. At $B_2$, the field-polarized state is reached. In the phase diagrams, the weak in-plane anisotropy is reflected. For the $[100]$ direction the AFM 4 is more stable for higher fields compared to the  $[110]$ direction, visible through the higher critical magnetic field $B_2$ in the magnetization.

\subsection{Eu(Rh$_{1-x}$Co$_{x}$)$_2$Si$_2$}
\subsubsection{Magnetization}
\begin{figure} [ht!]
    \centering
    \includegraphics[width=1\linewidth]{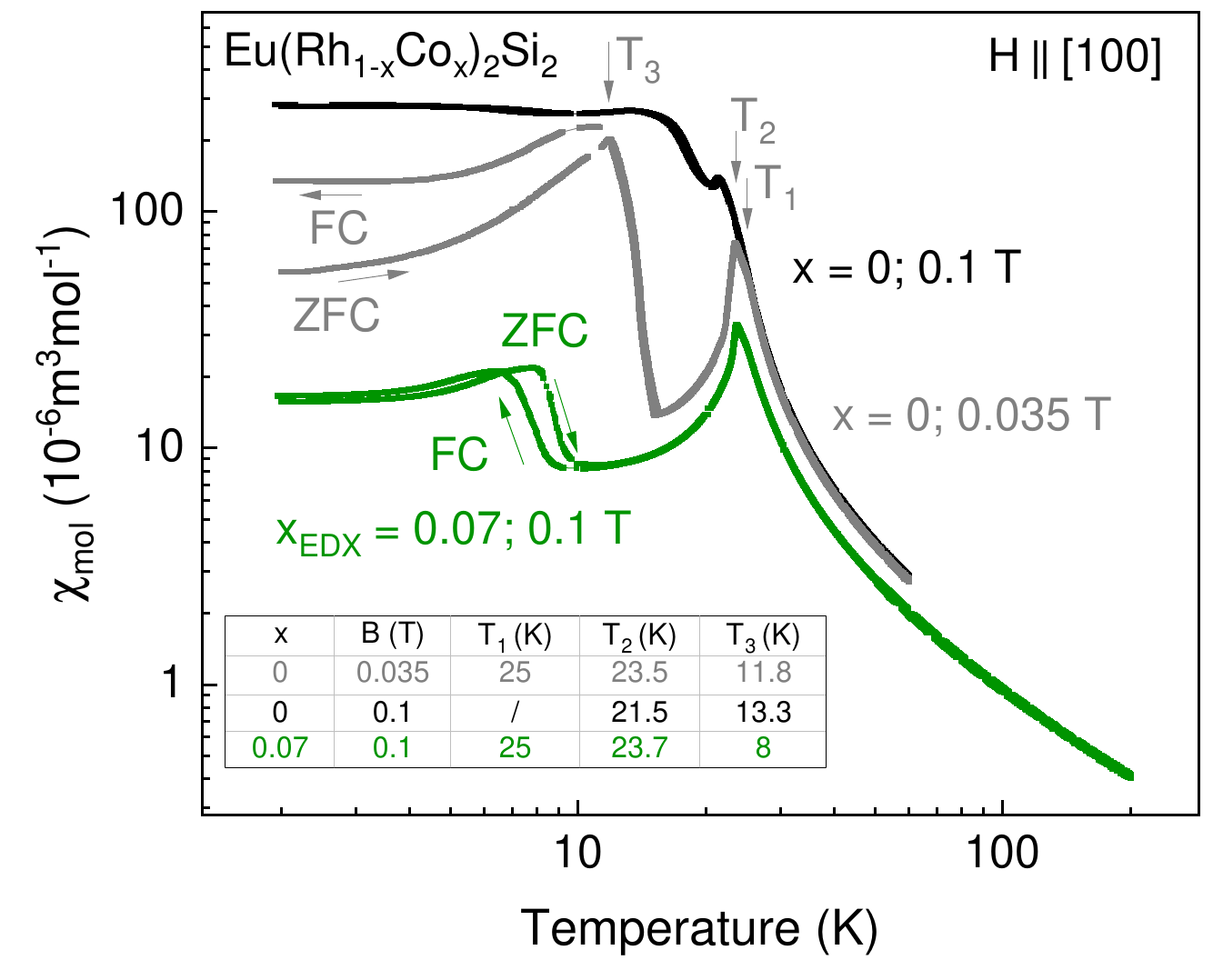}
    \caption{Eu(Rh$_{1-x}$Co$_{x}$)$_2$Si$_2$: Magnetic susceptibility as a function of temperature for a sample with $x$ = 0 at $\mu_0H=$ 0.035\,T (grey curve) and $\mu_0H=$ 0.1\,T (black curve) as well as for a sample with $x_{\rm EDX} = 0.07$ at $\mu_0H=$ 0.1\,T (green curve). The magnetic field was applied along the $[100]$ direction.} 
    \label{fig: MvT_0p07}
\end{figure}

\noindent Fig.~\ref{fig: MvT_0p07} shows the magnetic susceptibility as a function of temperature for the cobalt concentration  $x_{\rm EDX} = 0.07$ in comparison to  $x_{\rm EDX} = 0$, where the denoted substitution concentration was measured on the corresponding single crystal. The magnetic field was applied along the $[100]$ direction. For the unsubstituted compound EuRh$_2$Si$_2$ the magnetic transition at $T_1$ occurs up to a critical magnetic field of $\mu_0H = 0.035$\,T. For higher magnetic fields at 0.1\,T, the transition is suppressed and merges in $T_2$. The substituted system with $x_{\rm EDX} = 0.07$ still reveals antiferromagnetic ordering, and the three magnetic transitions are observed at low fields. In contrast to $x=0$, the transition at $T_1$ is now visible up to a critical magnetic field of $\mu_0H =  0.1$\,T, which leads to similar behavior of the 0.03\,T curve of the unsubstituted system and the 0.1\,T curve of the substituted sample with $x_{\rm EDX} = 0.07$. While the transition temperature $T_1$ remains constant under the influence of substitution, the AFM 1 phase is stabilized towards higher  magnetic fields.\\
\noindent In Fig.~\ref{fig:MvT_ACT_Vergleich}(a), the temperature-dependent magnetic susceptibility of an $x=0.08$ sample is compared with samples of higher Co concentration of $x_{\rm EDX} > 0.07$. The concentrations refer to the measured single crystals. For $x_{\rm EDX} = 0.08$ a steep increase in magnetic susceptibility with decreasing temperature is observed followed by a large drop.  Between the cooling and the heating curve, a huge thermal hysteresis with a width of $\sim 18$\,K appears, which is characteristic for a first-order phase transition. These signatures are typical in Eu compounds for a valence transition. The valence transition occurs at the valence transition temperature $T_{V} = 44$\,K, which is determined by the middle of the rise of the cooling curve. 
However, at $T_{\rm N} = 24$\,K a small kink and thus antiferromagnetic ordering is still present. Possible origin may be a distribution of different Co concentrations within the sample, where a part of the sample with a lower concentration $x_{\rm EDX}\leq 0.07$ is still in the antiferromagnetically ordered regime. Another reason could be that not all Eu ions contribute to the valence transition, as suggested in Ref.~\cite{mitsuda2020valence}, and thus some ions are still in the magnetically divalent state and give an antiferromagnetic contribution.\\
With increased Co concentrations $x_{\rm EDX} = 0.116$ and $x_{\rm EDX} = 0.119$, the valence transition is shifted to higher temperatures. The antiferromagnetic ordering is completely suppressed. The absence of magnetic ordering is also reflected in a reduced effective magnetic moment of $\mu_{\rm eff} = (5.8\,\pm 0.4)\mu_B$ for $x_{\rm EDX} = 0.119$ compared with the large effective magnetic moment of Eu${^{2+}}$ for $x = 0$ with $\mu_{\rm eff} = (7.76\,\pm 0.08)\mu_B$. 
The effective magnetic moment was obtained from a Curie-Weiss fit in the temperature range of 200\,K to 300\,K. For $x_{\rm EDX} = 0.116$ the valence transition occurs at $T_{V} = 84$\,K, and for $x_{\rm EDX} = 0.119$ at $T_{V} = 105$\,K. As expected from the general $p - T$ phase diagram following the first-order phase transition line, the hysteresis becomes smaller as the system gets closer to the valence crossover region. Between the $x_{\rm EDX} = 0.119$ and $x_{\rm EDX} = 0.116$ samples, slight changes in the Co substitution induce large differences in the valence transition temperature. Furthermore, tiny variations in the chemical composition can strongly affect the transition temperature as reported in \cite{kliemt2022influence} for the related compound EuPd${_2}$Si${_2}$. Since we observed the appearance of a homogeneity region for the 122 phase with minor differences in the Eu and Si concentrations, respectively, this could also cause the slight difference in $T_{V}$. With further increase of the substitution concentration, a moderate increase of the susceptibility at the valence crossover temperature $T_{V'} = 220$\,K for $x_{\rm EDX} = 0.228$ is observed, as shown in the inset of Fig.~\ref{fig:MvT_ACT_Vergleich}(a). In this case, there is no more thermal hysteresis. This suggests, that this concentration presents no  first-order valence transition anymore but instead is situated in the valence-crossover regime.

\subsubsection{Resistivity}
\begin{figure}[ht!]
    \centering
    \includegraphics[width=1\linewidth]{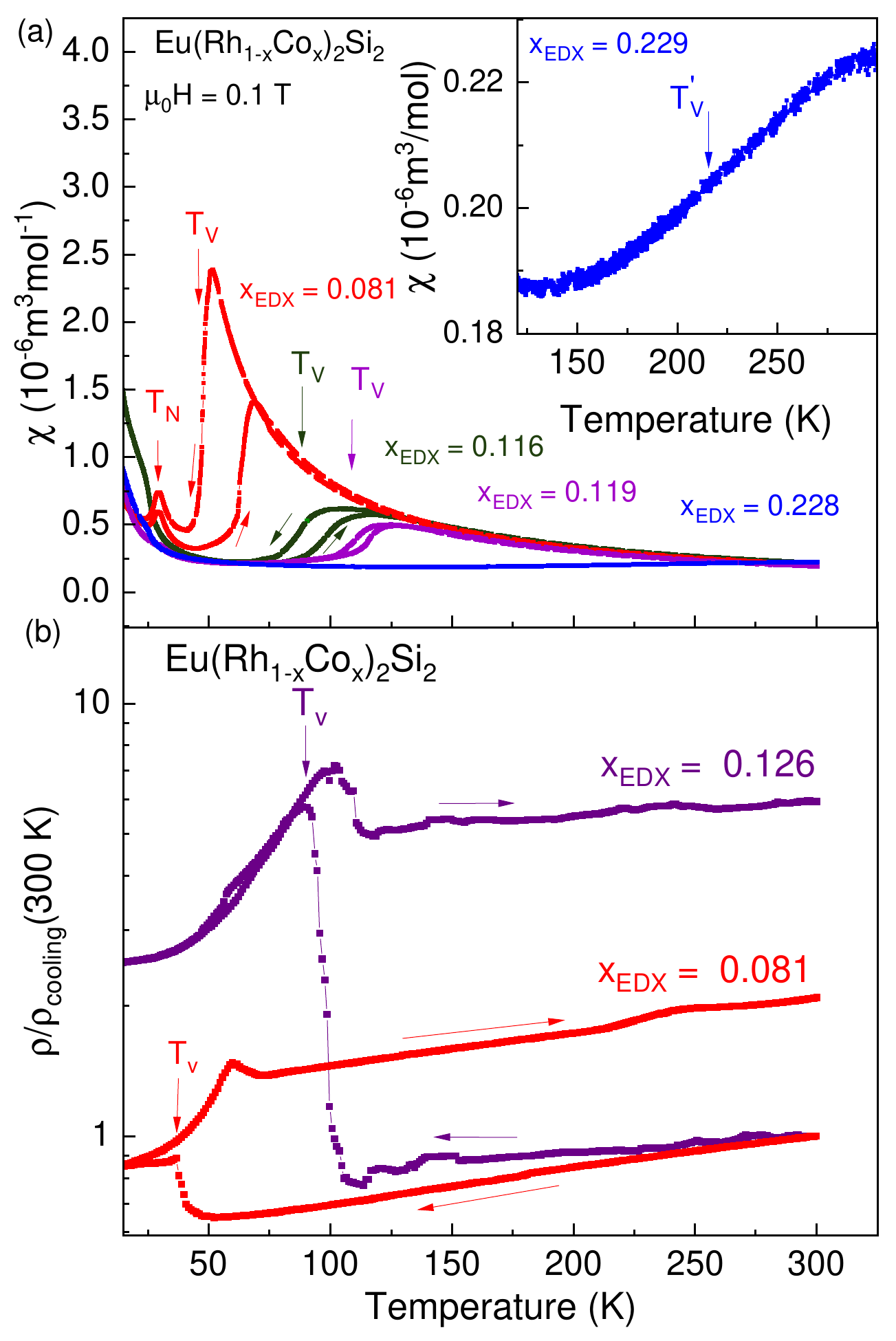}
    \caption{Eu(Rh$_{1-x}$Co$_{x}$)$_2$Si$_2$:(a) Magnetic susceptibility as a function of temperature for three different Co concentrations at an applied magnetic field of $B = 0.1$\,T. $T_V$ and $T_{V}$' denote the valence transition temperature and the valence crossover temperature, respectively. (b) Normalized temperature dependent electrical resistivity for two different Co concentrations.}
    \label{fig:MvT_ACT_Vergleich}
\end{figure}

\noindent To study the influence of Co substitution on the transport properties, temperature-dependent resistivity was measured for $x_{\rm EDX} = 0.126$ and $x_{\rm EDX} = 0.081$ (Fig.~\ref{fig:MvT_ACT_Vergleich}(b). The resistivity was normalized to the value of the resistivity $\rho (T= 300\,\rm K)$ of the cooling curve and the valence transition temperature was determined by the middle of the rise. 
For $x_{\rm EDX} = 0.126$ a sharp step appears at $T_V = 95$\,K with a large hysteresis of 15\,K, which is characteristic of a first-order phase transition. The resistivity increases by one order of magnitude with temperature. During one cooling cycle, the sample undergoes the valence transition, which is associated with a large compression of the volume cell. This possibly leads to microcracks in the sample, and thus the resistivity in the following heating cycle is a factor of six higher compared to the cooling cycle, similar to what was observed in  Eu(Rh$_{1-x}$Ir$_{x}$)$_2$Si$_2$  \cite{seiro2011stable}. For $x_{\rm EDX} = 0.081$ a sharp jump appears at $T_V = 37$\,K with a huge hysteresis of 23\,K and an overall similar $T$-dependence as described before for the sample with $x_{\rm EDX} = 0.126$. There is no anomaly of magnetic ordering at $T_{\rm N} = 25$\,K as observed in the magnetic susceptibility.

\subsubsection{Heat capacity}
\begin{figure}[ht!]
    \centering
    \includegraphics[width=1\linewidth]{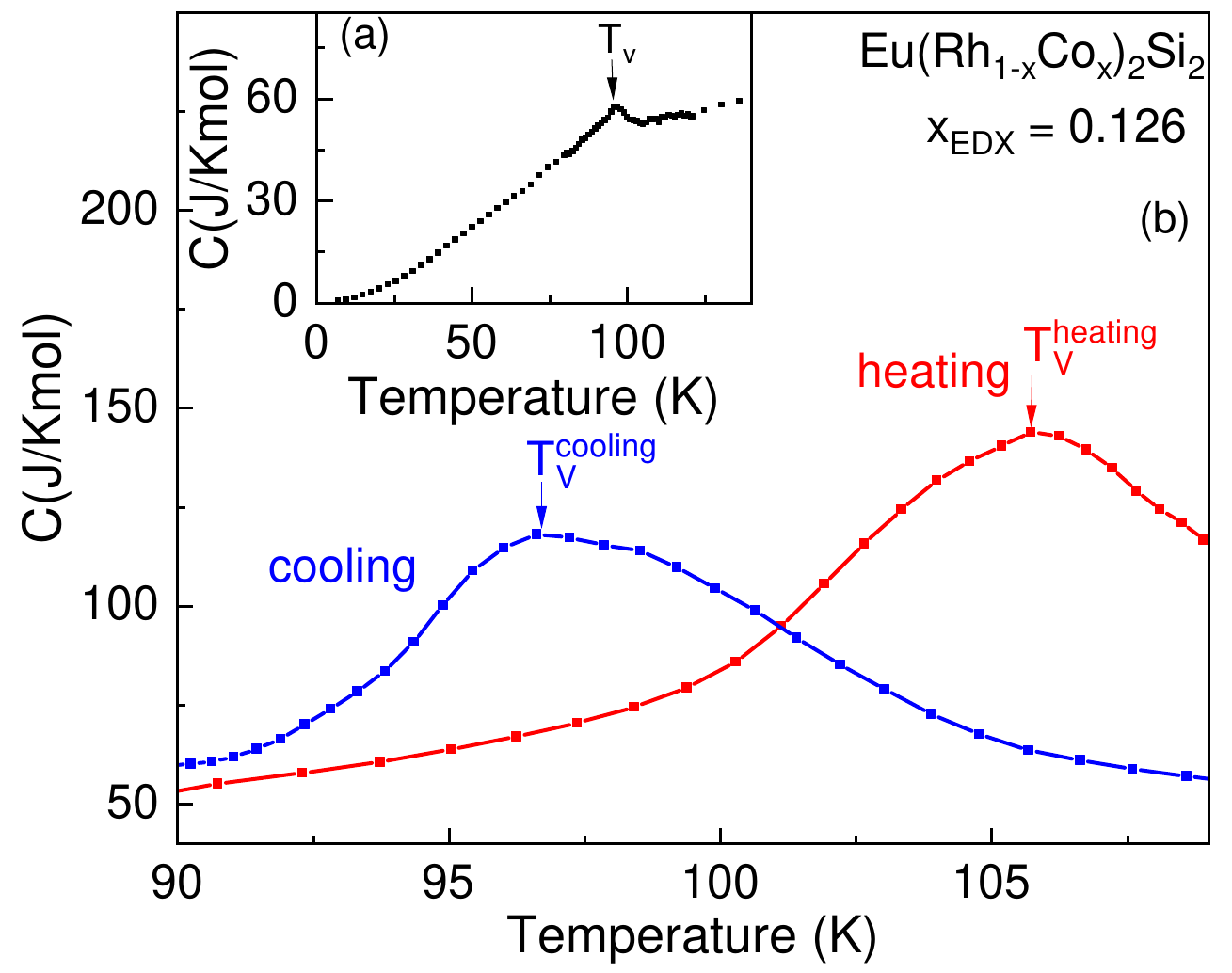}
    \caption{Eu(Rh$_{1-x}$Co$_{x}$)$_2$Si$_2$: (a) Heat capacity as a function of temperature for $x_{\rm EDX}$ = 0.126 . (b) Heat capacity as a function of temperature determined separately for cooling (blue curve) and heating (red curve) of long heat pulses through the valence transition.}
    \label{fig:HC_xEDX_0p126}
\end{figure}

\noindent In Fig.~\ref{fig:HC_xEDX_0p126}(a), the heat capacity as a function of temperature is shown for a sample with $x_{\rm EDX}$ = 0.126. In accordance with the resistivity data displayed in Fig.~\ref{fig:MvT_ACT_Vergleich}(b), the heat capacity shows an anomaly at the valence transition temperature $T_V$ = 95\,K. To investigate the first-order nature of the phase transition, the heat pulses were evaluated separately for the cooling (blue data) and heating (red data) process with single-slope analysis. 
A large difference of $\sim$ 10\,K is observed between the valence transition temperature for the coolingcurve,  $T^{\rm cooling}_V$ = 96\,K and the heating curve, $T^{\rm heating}_V$ = 106\,K. The hysteresis that appears in the heat capacity is a further indication of a first-order phase transition and to the best of our knowledge has not yet been observed in heat-capacity measurements of Eu compounds showing a valence transition.

\subsubsection{Phase diagram}

\begin{figure}[ht!]
    \centering
    \includegraphics[width=1\linewidth]{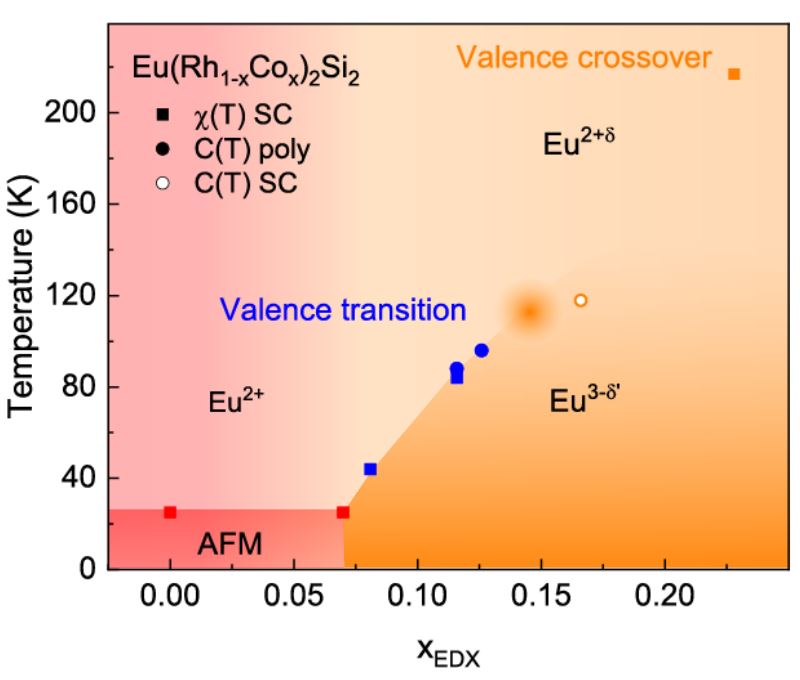}
    \caption{Temperature - substitution phase diagram for Eu(Rh$_{1-x}$Co$_{x}$)$_2$Si$_2$.}
    \label{fig:PD substitution}
\end{figure}

\noindent We summarize our results obtained from the magnetic susceptibility and heat capacity measurements in a  temperature-substitution phase diagram of Eu(Rh$_{1-x}$Co$_{x}$)$_2$Si$_2$ with $0.07\leq x_{\rm EDX} \leq 0.23$ in Fig.~\ref{fig:PD substitution}. 
For $x_{\rm EDX}\leq 0.07$, europium is still in the  divalent state and shows antiferromagnetic ordering below $T_{\rm N} = 24$\,K. Although a clear increase of the magnetic ordering temperature is not observed for $x_{\rm EDX} = 0.07$, as expected from the general Eu phase diagram \cite{onuki2020unique} and pressure experiments on EuRh$_2$Si$_2$ \cite{honda2016pressure}, an enhanced critical magnetic field of the AFM 1 phase is realized, which might indicate the stabilization of magnetic order. For increasing cobalt concentration, the system reveals a sharp valence transition for $x_{\rm EDX} = 0.081$, which is expected to occur between the magnetic divalent Eu$^{2+}$ and the valence-fluctuating  Eu$^{(3-\delta')+}$ state. 
This first-order phase transition is characterized by a pronounced thermal hysteresis in the magnetic susceptibility, the resistivity, as well as in the heat capacity. A further increase of the substitution concentration within the valence transition regime from $x_{\rm EDX}= 0.081$ to $x_{\rm EDX} = 0.126$ results in an almost linear increase of the valence transition temperature towards higher temperatures. 
This is accompanied by a decrease in the width of the first-order hysteresis, and thus shows that the critical endpoint is approached. For the  concentration level $x_{\rm EDX} = 0.166$, the system enters the valence crossover regime with the intermediate Eu$^{(2+\delta)+}$ state at high temperatures, which was observed in the heat-capacity data of a polycrystalline sample (not shown). Hence, the critical endpoint, where the first-order transition terminates followed by the  crossover region can be localized in the region 0.119 $<$ $x^{\rm CEP}_{\rm EDX}$ $<$ 0.166. With further increasing concentration, the valence crossover transition was clearly observed in the susceptibility of a single crystal with $x_{\rm EDX}$ = 0.23 at higher temperatures, where thermal hysteresis no longer occurs, indicating the crossover regime. In addition, this valence crossover takes place in a much broader temperature range of about 120\,K. Further increasing of the cobalt concentration may shift the valence crossover transition to room temperature. 
Compared to the substitution series Eu(Rh$_{1-x}$Ir$_{x}$)$_2$Si$_2$ \cite{seiro2011stable}, where the critical end point is located in the range of 0.5 $<$ $x$ $<$  0.75 and thus a strong influence of disorder is expected, in Eu(Rh$_{1-x}$Co$_{x}$)$_2$Si$_2$ the critical endpoint can be approached with a lower substituent concentration level. This opens the possibility to study in detail the properties around the CEP, with lower amount of disorder. \\ 
The valence transition/crossover caused by chemical pressure can now be compared to hydrostatic pressure applied on EuRh$_2$Si$_2$ single crystals \cite{honda2016pressure}. 
The degree of substitution can be converted to pressure using the formula $p_{\rm Co} = \Delta V/ V(x=0) \cdot K$, where $K$  denotes the bulk modulus. Here, a typical bulk modulus of $K$ = 100\,GPa \cite{onuki2020unique} is assumed. 
$\Delta V$/ $V$ indicates the change in unit cell volume with respect to unit-cell volume $V$ for $x = 0$ and can be taken from the PXRD data shown in  Fig.~\ref{diffractogram}(c). Thus, for chemical induced pressures below $p_{\rm Co}\leq 0.7$\,GPa the system orders antiferromagnetically, which is comparable but slightly lower than reported in the work of Honda et al. \cite{honda2016pressure}, where the antiferromagnetic phase is stable up to $p_{\rm AFM}\leq$ 0.96\,GPa. For higher substitution levels, AFM ordering disappears and a first-order phase transition is realized between 1.2 $\leq p_{\rm Co} \leq$ 1.7\,GPa. Comparing that to measurements under pressure, the valence transition occurs between 0.96 $\leq p \leq$ 2\,GPa. From substitution experiments, the critical endpoint would be estimated to be at $p^{\rm CEP}_{\rm Co}$ = 1.7\,GPa, which is slightly lower in comparison to $p^{\rm CEP}$ evaluated in the hydrostatic pressure experiments \cite{honda2016pressure}. The overall agreement between hydrostatic pressure experiments on pure EuRh$_2$Si$_2$ crystals and the here presented substitution series Eu(Rh$_{1-x}$Co$_{x}$)$_2$Si$_2$ is remarkable and shows, that the introduced disorder through cobalt substitution has only minor effects on the overall phase diagram. 
\section{Summary}

\noindent In conclusion, we have succeeded in growing single crystals of the substitution series Eu(Rh$_{1-x}$Co$_{x}$)$_2$Si$_2$ using Eu-flux. The chemical analysis revealed slight deviations from the ideal 1:2:2 stoichiometry with respect to Eu and Si, but with constant (Rh+Co) substitution level within a sample. For EuRh${_2}$Si${_2}$, we have found a weak magnetic anisotropy within the basal plane, which is reflected in different $B-T$ phase diagrams for the $[100]$ and $[110]$ directions. These phase diagrams reveal a complex magnetic behavior and four different AFM phases could be clearly distinguished, while we suspect a fifth magnetic phase at low temperatures and small magnetic field representing the magnetic ground state. 
Based on these observations, for the substituted system with $x_{\rm EDX} = 0.07$ we observed an enhanced critical magnetic field of the AFM 1 phase, stabilized  by the Co-substitution. 
Within the concentration range 0.08 $\leq$ $x_{\rm EDX}$ $\leq$ 0.119, the system undergoes a first-order phase transition, where a valence transition from the magnetic divalent to the intermediate Eu$^{(3-\delta')+}$ is realized, which is accompanied by the disappearance of the antiferromagnetic ordering. 
The first-order nature was revealed in magnetic susceptibility, resistivity, and heat capacity by a sharp and large hysteresis around the valence transition temperatures $T_V$. With increasing concentration level $x_{\rm EDX} \geq$ 0.15, the system changes to the valence crossover regime and  no thermal hysteresis was observed at $T_V'$ = 220\,K. As a result, we locate the critical endpoint in the vicinity of 0.119 $<x_{\rm EDX}<$ 0.166. 
Therefore, the substitution level to approach the critical endpoint is rather low, which offers the possibility to study in detail the CEP under ambient or small hydrostatic pressures without sizeable amount of disorder. To this end, the  presented substitution series Eu(Rh$_{1-x}$Co$_{x}$)$_2$Si$_2$ is a suitable system to detect critical elasticity in an intermetallic compound.

\begin{acknowledgments}

We thank Christoph Geibel for valuable discussions and T. F\"orster for technical support. We acknowledge funding by the Wilhelm and Else Heraeus Foundation
and the Deutsche Forschungsgemeinschaft (DFG, German Research Foundation) via the TRR 288 (422213477, project A03). 
\end{acknowledgments}

\section{References}

\bibliography{EuRh2Si2}

\end{document}